\documentclass[aps,twocolumn,groupedaddress,showpacs]{revtex4-1}
\usepackage{epstopdf}
\usepackage{physics}
\usepackage{tikz,lipsum,lmodern}
\usepackage[export]{adjustbox}
\usepackage{subcaption}
\usepackage{amsmath}
\usepackage{amssymb}
\usepackage{svg}
\usepackage{xcolor}
\usepackage{bbold}
\usepackage{physics}
\usepackage[unicode=true,pdfusetitle,
 bookmarks=true,bookmarksnumbered=false,bookmarksopen=false,
 breaklinks=false,pdfborder={0 0 0},backref=false,colorlinks=true,citecolor=blue,urlcolor=violet 
]{hyperref}
\usepackage{xcolor}
\usepackage{graphicx}
\newcommand{\orcid}[1]{\href{https://orcid.org/#1}{\includegraphics[width=10pt]{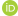}}}
\usepackage{setspace}
\usepackage{soul}

\newcommand{\suc}{\mathrm{succ}}
\begin{document}
\bibliographystyle{apsrev}
\title{Semi-device-independent certification of quantum non-Markovianity using sequential Random Access Codes}
 \author{Abhinash Kumar Roy$^{1,2,6}$\orcid{0000-0001-7156-1989}}
 \email{abhinash.roy@students.mq.edu.au}
\author{Varun Srivastava$^{1,2,6}$\orcid{0000-0002-3907-5304}}
\email{varun.srivastava@students.mq.edu.au}
\author{Soumik Mahanti$^{3,4}$\orcid{0000-0002-0380-0324}}
\email{soumikmh1998@gmail.com}
\author{Christina Giarmatzi$^{5,6}$\orcid{0000-0001-8254-2169}}
\email{christina.giar@gmail.com}
\author{Alexei Gilchrist$^{1,2,6}$\orcid{0000-0003-0075-5174}}
\email{alexei@entropy.energy}
\affiliation{$^1$Department of Physical and Mathematical Sciences, Macquarie University, Sydney NSW, Australia}
 \affiliation{$^2$Centre for Engineered Quantum Systems, Macquarie University, Sydney NSW, Australia}
 \affiliation{$^3$S.N. Bose National Center for Basic Sciences, Block JD, Sector III, Salt Lake, Kolkata 700106, India}
 \affiliation{$^4$Indian Institute of Science Education and Research Kolkata, Mohanpur, 741246 West Bengal, India}
 \affiliation{$^5$School of Computer Science, University of Technology Sydney,
Ultimo, Sydney, New South Wales 2007, Australia}
\affiliation{$^6$ARC Centre of Excellence for Engineered Quantum Systems,
St. Lucia, Brisbane, Queensland 4072, Australia}
\begin{abstract}
The characterization of multi-time correlations in open quantum systems is of fundamental importance. In this work, we investigate multi-time processes using the process matrix formalism and show that the presence of a quantum non-Markovian environment plays a significant role in enhancing the communication capacity in sequential prepare-transform-measure Quantum Random Access Codes (QRAC). The correlated environment enables a quantum advantage to multiple parties, even with projective measurements. In particular, we show that the Markovian and classical non-Markovian processes, i.e. quantum processes with classical feedback from the environment, do not yield sequential quantum advantage. In contrast, it is possible to achieve an advantage in the presence of a quantum non-Markovian environment. Therefore this approach allows a semi-device-independent certification of quantum non-Markovianity. As opposed to entanglement-detection criteria which require the knowledge of the complete process, this method allows to certify the presence of a quantum non-Markovian environment from the observed measurement statistics. Moreover, quantum memory ameliorates the unambiguous certifiable region of unsharp instruments in a semi-device-independent manner.

\end{abstract}

\maketitle

\section{Introduction}
Realistic quantum systems are not isolated and are subject to decoherence due to presence of an environment \cite{Breuer_Book,Schlosshauer_2019}. To develop error mitigation protocols, noise characterisation in multi-time processes is of fundamental importance. Current error correction techniques rely on the assumption that the noise across different time steps is uncorrelated, i.e., the process is Markovian \cite{White2020,PhysRevLett.121.060401,PhysRevLett.121.220502,Sarovar2020detectingcrosstalk}. However, it turns out that non-Markovianity is the norm rather than an exception and correlated noise has been identified in the state-of-the-art quantum devices of IBM and Google \cite{giarmatzi_2023,Modi_2022_IBM,2021_Superconducting}.

Characterization of non-Markovianity in a quantum scenario is a non-trivial task, with past approaches lacking a necessary and sufficient condition \cite{2018_Li_Physics_reports,RMP_2016_Breur,RMP_2017_Vega,Rivas_2014,Shrikant_Frontiers,Li_2019_EPL_Review}. Recently, an operational approach to quantum non-Markovianity was developed using process matrices (also known as process tensors/quantum combs for the multi-time causally ordered scenario) \cite{Costa_2016,PhysRevLett.120.040405,Modi_PRA_2018,PRXQuantum_Modi_2021}. The formalism uses the concept of causal-break interventions at each time step to check for conditional dependence of statistics of the future evolution of the system on previous time steps. A Markovian process has a specific structure --- the process matrix is a tensor product of channels connecting the labs \cite{Costa_2016,Giarmatzi_2021}. This characterization was recently used to divide the set of non-Markovian processes in two subsets: classical non-Markovian processes and quantum non-Markovian processes \cite{Giarmatzi_2021}. In the former subclass, the process can be simulated with a classical feedback mechanism, whereas in the latter, such simulation is not possible due to coherent correlation in noise across the time steps. This division of non-Markovian processes is motivated by the fact that for classical correlated noise, it is easier to identify the source and mitigate it compared to coherently correlated noise \cite{Xu_2013_NatComm,Bordone_2012,PRA_2012_class_quantum_mem}. The study of this division is of both fundamental and practical importance.

Recently, it was shown that entanglement across the relevant bi-partition of a process matrix corresponds to the presence of coherently correlated noise~\cite{Giarmatzi_2021}. However, this is a sufficient but not necessary criterion for certification of quantum non-Markovianity~\cite{Nery_2021_Quantum}. In addition, the certification relies on knowing or having a good guess for the process matrix, which is obtained through a resource-expensive process matrix tomography~\cite{giarmatzi_2023,PRA_Goswami_2021}. In this work, we provide an alternative criterion for certification of quantum non-Markovianity that relies only on the joint statistics with minor assumptions on the dimensions of input-output Hilbert spaces and the sharpness of measurements performed at the intermediate time steps. It therefore provides a \emph{semi-device independent} certification of quantum non-Markovianity. This form of certification of quantum non-Markovian environment is possible without trusting the preparation and measurement devices in the labs, making it a stronger certification criterion as opposed to one that requires process tomography, where trust in instruments is essential. Device-independent certification techniques have been used widely. For example in secure Quantum Key Distributions~\cite{Acin2007PRL,pironio2009NJP}, randomness certification~\cite{pironio2010nature,colbeck2012NaturePhys}, dimension witnessing~\cite{Gallego2010PRL,Brunner2013PRL}, and communication complexity~\cite{buhrman2016PNAS,Buhrman2010RevModPhys}, which relies on the violation of some form of Bell's inequality~\cite{PhysicsPhysiqueFizika.1.195}. This enables a certification scheme from nonlocal quantum correlation without considering anything about the internal functioning of the preparation or measurement apparatus. Assuming an upper bound on the Hilbert space dimension, the certification schemes have been extended in a semi-device-independent manner to prepare-measure scenarios \cite{Pawlowski2011PRA,Li2011PRA,Bowles2014PRL,Tavakoli2018PRA,Miklin2020PRR,Mohan_2019,tavakoli2020Science_adv, Mukherjee2021semi, miklin2021Quantum,Mukherjee2023PRA}, where the quantum advantage has been shown to violate the preparation-noncontextual~\cite{pan2019Scientific_Reports,Pan2021PRA,Mukherjee2023PRA} or Kochen-Specker noncontextual bound~\cite{Gupta2023PRL}. 

Here, we use a three-party sequential ($2\!\rightarrow\!1$) Quantum Random Access Code (QRAC) game \cite{Mohan_2019,Miklin2020PRR,Foleto2020Experimental,PhysRevLett.125.080403} with rank-one projective (sharp) measurements at the intermediate party, to find a bound on the success probability of the third party which is satisfied by all Markovian and classical non-Markovian processes. Due to the fact that valid process matrices obey linear constraints \cite{Oreshkov_2012_Natcomm,Araújo_2015_NJP} and that the success probability of a QRAC game is a linear function of the process matrix, we cast our problem of finding processes that yield sequential advantage as a Semi-Definite Program (SDP) that can be solved efficiently \cite{boyd_vandenberghe_2004,Daniel_book_IOP_2022}. 
We show that there exist multi-time processes yielding sequential advantage and that a higher success probability at the third party is associated with a larger quantum correlation across time steps. Therefore the technique provides a measure of quantum non-Markovianity through the statistics of measurement results in a semi-device-independent manner. Moreover, we relax the sharp measurement constraint in the intermediate lab and perform a robustness analysis for the certification. In particular, we show that assuming a lower bound on the unsharpness parameter, it is still possible to certify the presence of a quantum non-Markovian environment. In addition, we find a region to unambiguously certify unsharp instruments at the intermediate lab. To our knowledge, this is the first approach that provides a semi-device-independent certification of a class of system-environment interactions, in particular, processes with coherently correlated noise across the time steps.

Our work is organized as follows: In Section \ref{sec2}, we briefly review the process matrix formalism and the characterization of Markovian and non-Markovian processes (and their subclasses) within this formalism. In  Section \ref{sec3}, we review standard prepare-measure QRAC games in two-party and sequential scenarios. In Section \ref{sec4}, we provide our results on semi-device-independent certification of quantum non-Markovianity. First, we provide a bound on the success probability in the sequential QRAC for Markovian (\ref{RACMark}) and classical-memory processes (\ref{SeqRACclnonMark}) and provide a feasibility problem to search for quantum memory processes violating the bound (\ref{Feasibilityprob}). Then we show the existence of processes yielding sequential advantage by using an SDP (\ref{processmatfeasible}). We discuss the relationship between the amount of quantum non-Markovianity and the sequential quantum advantage in (\ref{EnatnglProcess}). We provide robustness analysis for certifying quantum memory in (\ref{subsec:Robust}). In (\ref{Unambiguous}) we improve the conditions required to unambiguously certify unsharp instruments and provide a tighter bound.
In Section \ref{sec5} we discuss how our results compare with prior works and discuss various extensions.

\section{Preliminaries}\label{sec2}
\subsection{Process Matrix Formalism}

The process matrix formalism is a natural way of studying multi-time processes \cite{Giarmatzi_2021,White2020,PhysRevLett.120.040405,Fabio_2018_npj,Fabio_2018_PRA}. It captures the most general type of quantum processes that involve the evolution of a system and intervening parties at each time step, that measure and transform the system, to analyse a quantum process. The process matrix gives us a prescription on acquiring the joint probability distribution of results of each party by a generalisation of Born's rule \cite{Shrapnel_2018_NJP}.

Each party is represented by a \textit{quantum instrument} $\{\mathcal{M}^{A}_{i|x}\}$, where $\mathcal{M}^{A}_{i|x}$ a completely positive (CP) trace non-increasing map associated with each outcome $i$ of the party and their sum is a completely positive trace preserving (CPTP) map for all setting $x$, $\mathcal{M}^{A}_{x}=\sum_{i}\mathcal{M}_{i|x}^{A}$. The joint probability distribution is then given by the generalised Born's rule,
\begin{equation}
    \begin{aligned} 
     P(i,j,k,
     \cdot\cdot\cdot| x,y,z,\cdot\cdot\cdot)=\qquad\qquad\qquad\qquad\qquad&\\
     \operatorname{Tr}\left[W^T \left(M^{A_1 A_2}_{i|x} \otimes M^{B_1 B_2}_{j|y} \otimes M^{C_1 C_2}_{k|z} \otimes...\right)\right],    
    \end{aligned}
\end{equation}
where $M^{A_1 A_2}_{i|x}$ is the Choi-Jamio\l kowski (CJ) matrix \cite{JAMIOLKOWSKI1972275,CHOI1975285} for the map $\mathcal{M}^{A}_{i|x}$ and $W \in \mathcal{L}(\mathcal{H}^{A_1} \otimes \mathcal{H}^{A_2} \otimes \mathcal{H}^{B_1} \otimes \mathcal{H}^{B_2} \otimes \mathcal{H}^{C_1} \otimes \mathcal{H}^{C_2} \otimes ...)$ is known as the \textit{process matrix}. Note that we have chosen to associate a transpose with the $W$ matrix as compared with some other formulations, e.g. \cite{Giarmatzi_2021}. The CJ matrix $M^{A_1 A_2}_{i|x} \in  \mathcal{L}\left(\mathcal{H}^{A_1} \otimes \mathcal{H}^{A_2} \right)$ is defined as:
\begin{equation}\label{choi}
    \begin{aligned}
        M^{A_1 A_2}_{i|x}=&
        \mathcal{I}\otimes \mathcal{M}^{A}_{i|x} \left(| \mathbb{1} \rangle \langle \mathbb{1}| \right),
    \end{aligned}
\end{equation}
where $\mathcal{I}$ is the identity map, $\mathcal{M}^{A}_{i|x}: \mathcal{L}\left(\mathcal{H}^{A_1}\right) \rightarrow \mathcal{L}\left(\mathcal{H}^{A_2}\right)$ and $|\mathbb{1}\rangle = \sum_{i=1}^{d_{A_1}} |i\rangle|i\rangle \in \mathcal{H}^{A_1} \otimes \mathcal{H}^{A_1}$ is the un-normalised maximally entangled state with orthonormal basis $\{ |i\rangle\} \in \mathcal{H}^{A_1}$ and $i=1,2,3...,d_{A_1}$.

The process matrix $W$ captures the information of everything except the intervening parties in a quantum process and hence makes it a useful tool to study muti-time correlations in non-Markovian quantum processes.

\subsection{Markovian and non-Markovian Processes}
While there are various characterisations of Markovian and non-Markovian quantum processes in the literature, we take an operational approach discussed in \cite{Giarmatzi_2021} using the process matrix $W$, which is the general representation of a multi-time quantum process. 

\subsubsection{Markovian processes}
For a classical stochastic process, multi-time correlations for a random variable $X_{t}$ at time $t$ are given through a joint probability distribution $P(X_{t},X_{t-1},...,X_{0})$ where $X_{i}$ is the value of random variable at that time step. A Markovian process then has the following property:
\begin{equation}
    P(X_{k} | X_{k-1}, X_{k-2},...,X_{0})=P({X_{k}|X_{k-1}})
\end{equation}
for all time steps. For a multi-time quantum process the process matrix that represents Markovian processes has the following form \cite{Costa_2016,Shrapnel_2018_NJP, Giarmatzi_2021}
\begin{equation}\label{Wmark}
    W_{M}=\rho^{A_1} \otimes [\mathcal{D}]^{A_2 B_1} \otimes [\mathcal{D}]^{B_2 C_1} \otimes ...,
\end{equation}
where $\rho^{A_1}$ is the density matrix at the input of first lab and $[\mathcal{D}]^{A_2 B_1}$ (and all other terms), are the CJ-matrix of  CPTP  maps that satisfy the condition $\operatorname{Tr}_{B_1} \left( [\mathcal{D}]^{A_2 B_1}\right)= \mathbb{1}^{A_2}$. The output from the last party is discarded so the process matrix has an identity at the end. Any process matrix of this form represents a Markovian quantum process.

\subsubsection{Non-Markovian processes}
Process matrices that cannot be written in the form above represent a non-Markovian quantum process and can be further characterised by their memory being classical or quantum \cite{Giarmatzi_2021}.

\textit{Classical memory processes} are processes in which environment can be simulated using classical feedback mechanism which correlates the noise. In a general classical memory process the environment can be thought to measure the system at each time step, say between $t_j$ and $t_{j+1}$ and acquire a classical outcome $a_j$. The evolution of the system after this time can depend on $a_j$ but also on the information stored by the environment up to that point, noted by a classical variable $x_j$. The most general evolution of this type is represented by a CP map $\mathcal{T}^j_{a_j|x_j}:{X^j_O \rightarrow Y^{j+1}_I}$ from party $X$ to $Y$, where the sum the sum $\sum_{a_j}\mathcal{T}^j_{a_j|x_j}$ must be a CPTP map, i.e. each CP map has to be an instrument. In the CJ representation, an instrument satisfies the following

\begin{equation}
[\mathcal{T}_{a_j|x_j}]^{A^j_O A^{j+1}_I}\geq 0, \qquad \operatorname{Tr}_{A^{j+1}_I} \sum_{a_j} [\mathcal{T}_{a_j|x_j}]^{A^j_O A^{j+1}_I} = \mathbb{1}^{A^j_O}.
\end{equation} 

Finally, the classical information stored in the environment is discarded and the process matrix of a classical memory process is 

\begin{equation} 
W^{A^1\dots A^n}_{\textrm{CM}} = \sum_{\vec{x} \vec{a}} \bigotimes_{j=0}^{n-1} [\mathcal{T}_{a_j|x_j}]^{A^j_O A^{j+1}_I} P(x_j|\vec{a}_{|j},\vec{x}_{|j}), 
\label{classicallong}
\end{equation}

where $\vec{a}_{|j}\equiv\{a_{0},a_{1},...a_{j-1}\}$ and $\vec{x}_{|j}\equiv \{x_{0},x_{1},...x_{j-1}\}$ and $P(x_j|\vec{a}_{|j},\vec{x}_{|j})$ are the conditional probability of variable $x_{j}$. We also note that $\{[\mathcal{T}_{a_j|x_j}]^{A_{I}^{1}}\}_{a_{0}} \equiv \{\rho_{a_{o}|x_{0}}^{A_{I}^{1}}\}_{a_{0}}$ where $\{\rho^{A_{I}^{1}}_{a_{0}|x_{0}}\}$ are sub-normalised states and $P(x_0|\vec{a}_{|0},\vec{x}_{|0})=P(x_{0})$ is the marginal probability of initial variable $x_{0}$. A process matrix has classical memory if and only if it can be written in this way. For a more detailed discussion on classical memory processes we refer the reader to \cite{Giarmatzi_2021}.

\textit{Quantum memory processes} are the remaining set of process matrices, i.e $W_{QM} \in W/W_{CM}$. Owing to the fact that a process matrix maps a multi-time process into a multipartite state $W$, one can characterise a subset of quantum memory processes by checking for entanglement on the state \cite{Giarmatzi_2021}. This method provides a powerful tool for checking for quantum memory in a non-Markovian process. However, entanglement of the process can either be verified through process tomography, or through entanglement witnesses, which requires a good guess of the process.

\section{Random Access Code} \label{sec3}
Random access codes are a special class of communication complexity tasks where a sender has to encode a message in fewer bits than the original message and the receiver has to retrieve any subset of the message with highest success rate. One can thus think about optimization over different strategies to find the maximum probability to retrieve the required message. The success in RAC games is considered mostly in two types of scenarios, namely - the average success probability and the worst-case success probability. From here on, we only talk about the average success probability, whenever we mention success probability. It has been found that the quantum version of the game where the message is encoded in qubits instead of bits yields an advantage in the success probability to retrieve the message~\cite{ambainis1999dense,ambainis2008quantum}. In a typical Quantum Random Access Code (QRAC) scenario, Alice prepares a set of quantum states depending on random input messages and sends them to Bob. Bob performs a measurement on received states depending on the random subset of the message to be retrieved. The challenge and interest in QRACs arise from optimising the choice of quantum states and measurements to achieve specific objectives, like minimising the error rate or maximizing the amount of information transferred. There are several settings where one can study QRAC, such as preparation-measurement~\cite{ambainis1999dense}, preparation-transformation-measurement~\cite{Mohan_2019}, entanglement assisted~\cite{Pawlowski2010entanglement,Kanjilal2023PRA}, with shared randomness~\cite{ambainis2008quantum}, having parity-oblivious constraint \cite{chailloux2016optimal} etc. Here, we are interested in a preparation-transformation-measurement QRAC game in a ($2\!\rightarrow\!1$) setting, i.e. a 2-bit message has to be encoded in a qubit.

\subsection{Two party QRAC}
\begin{figure}[h]
    \includegraphics[width=0.9\columnwidth]{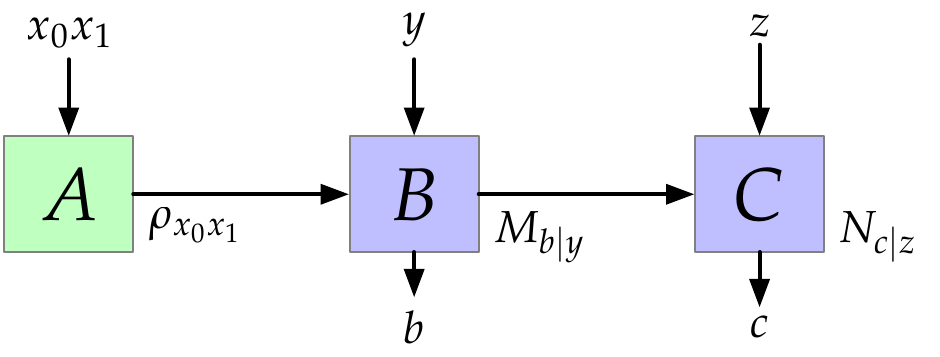}
    \caption{A sequential ($2\!\rightarrow\!1$) quantum RAC game. Alice prepares one of 4 states depending on the bit string she receives $x\in\{00,01,10,11\}$, Bob and Charlie try to guess the random bit $y\in\{0,1\}$ and $z\in\{0,1\}$ in the string respectively.} 
    \label{fig:2pRAC}
\end{figure}

Consider a two-party ($2\!\rightarrow\!1$) QRAC protocol. Alice randomly obtains a two-bit input message $x = x_0x_1$, where $x_i\in \{0, 1\}$, prepares a state $\rho_{x_0x_1}$ and sends it to Bob. Bob is given a random bit $y$, and his task is to guess the $y^{th}$ bit of Alice (see Fig.~\ref{fig:2pRAC}). They win the game if Bob's guess is right. In QRAC, Bob applies an instrument, $\{M_{b|y}\}$, on the received system depending on his input $y\in\{0,1\}$ where $\sum_{b}M_{b|y} =1 ~\forall~y$; and $b$ is a binary outcome corresponding to his guess. They win the game when $b = x_{y}$. Therefore the average success probability is given by
\begin{equation} \label{racpsucc}
    p_{\suc} = \frac{1}{8}\sum_{x,y}p(b = x_{y}|x,y),
\end{equation}
where the normalization factor arises from inputs $x_0, x_1$ and $y$ being equally probable. Now, if Alice is constrained to send only a classical bit, the optimal success probability is $3/4$. This can be easily seen through the following strategy: Alice always encodes one of her input bits, say the first one, and sends it to Bob. In this scenario, whenever $y=0$, Bob can correctly guess Alice's first bit, and whenever $y=1$, the guess will be random, i.e., $p(b\!=\!x_{0}|y\!=\!0) = 1$ and $p(b\!=\!x_{1}|y\!=\!1) = 1/2$, which leads to a success probability of $3/4$. Interestingly, if we allow Alice to encode her input in a qubit, there exists state preparations for Alice and measurements for Bob which can beat the classical optimal success probability. 

If Alice encodes her message in a quantum state $\rho_x$, the success probability is given by
\begin{equation}\label{qracpsucc}
    p_{\suc} = \frac{1}{8}\sum_{x,y}\operatorname{Tr}\left[\rho_{x}M_{x_{y}|y}\right].
\end{equation}
If we assume that Alice uses the following state assignments for the bit string, 
\begin{equation}\label{stateprep}
    \begin{aligned}
        \rho_{x_{0}x_{1}} = \frac{1}{2}\left(\mathbb{1}+\frac{1}{\sqrt{2}}\left((-1)^{x_{0}}\sigma_{z}+(-1)^{x_{1}}\sigma_{x} \right)\right),
    \end{aligned}
\end{equation}
and that Bob performs the following projective rank-1 measurements,
\begin{equation}\label{proj1}
\begin{aligned}
    M_{b|y} = \frac{1}{2}\left(\mathbb{I} + \frac{(-1)^{b}}{2}\left((1-(-1)^{y})\sigma_{x}+(1+(-1)^{y})\sigma_{z}\right)\right),
    \end{aligned}
\end{equation}
the optimal success probability is $p_{\suc} = (2+\sqrt{2})/4 >3/4$. This optimal success probability in prepare-measure QRAC self-tests that the prepared states form a square on a great circle of the Bloch sphere and the measurement operators are projectors on the diagonals \cite{Tavakoli2018PRA}.  

\subsection{Sequential QRAC}
In a sequential QRAC game, after performing the desired instrument on the state, Bob relays the post-measurement state to another party, say Charlie, who has to decode another independent subset of the message (see Fig.\ref{fig:2pRAC}). Alice prepares the state $\rho_{x_0x_1}$ and sends it to Bob who, depending on a random bit $y$, applies the instrument $\mathcal{M}_{y} = \{M_{b|y}\}$. The post-measurement state $\tilde{\rho}_{x_0x_1}$ is 
\begin{equation}
    \tilde{\rho}_{x_0x_1} = \frac{1}{2}\left[\mathcal{M}_{0}(\rho_{x_0x_1})+\mathcal{M}_{1}(\rho_{x_0x_1})\right],
\end{equation}
which is relayed to Charlie. Now, depending on a random bit $z$, Charlie performs an instrument $\mathcal{N}_{z} = \{N_{c|z}\}$. We are interested in the case where both Bob and Charlie beat the classical optimal success probability, which is $\frac{3}{4}$ irrespective of the number of independent sequential observers. In sequential QRAC, there is a trade-off between the extracted information and measurement-induced disturbance in the post-measurement state at lab B. If Bob performs the projective measurement as in Eq.~\eqref{proj1} on Alice's preparation given in Eq.~\eqref{stateprep}, and the resulting post-measurement states are sent to $C$, in the noiseless scenario the optimal success probability for Charlie is $(4+\sqrt{2})/8$.\\
In the process matrix formalism, the statistics obtained from three parties $A$, $B$  and $C$ are given by
\begin{equation}\label{eq:Wabc}
    p(a,b,c|x,y,z) = \operatorname{Tr}\left[W^T\left(M_{a|x}^{A_{1}A_{2}}\otimes M_{b|y}^{B_{1}B_{2}}\otimes M_{c|z}^{C_{1}C_{2}}\right)\right], 
\end{equation}
where, $W$ is the process matrix, and terms of the form $M^{P_{1}P_{2}}$ are the CJ-matrices of the operations performed by party $P$, with $P_{1}$ and $P_{2}$ representing the input and output Hilbert spaces respectively. 
Party $A$ performs a deterministic state preparation conditional on the two-bit setting she receives, $x=x_{0}x_{1}$. The CJ-matrix for the operation in her lab is
\begin{equation}\label{eq:op_A}
    M_{x}^{A_{1}A_{2}} =  \mathbb{1}^{A_{1}}\otimes\rho_{x}^{A_{2}},
\end{equation}
which is equivalent to discarding the state on $A_1$ and preparing a state deterministically on $A_2$, which is sent to $B$. With this specialisation, Eq.~\eqref{eq:Wabc} becomes
\begin{equation}\label{eq:Wbc}
    p(b,c|x,y,z) = \operatorname{Tr}\left[W^T\left(\mathbb{1^{A_1}}\otimes\rho_{x}^{A_2}\otimes M_{b|y}^{B_{1}B_{2}}\otimes M_{c|z}^{C_{1}C_{2}}\right)\right].
\end{equation}
Party $B$ performs a measurement and sends the post-measurement state to $C$. If the measurement operator corresponding to the setting $y$ and outcome $b$ is $E_{b|y}$, the CJ-matrix of $B$ is
\begin{equation}\label{eq:op_B}
    M_{b|y}^{B_{1}B_{2}} = |E_{b|y}\rangle\langle E_{b|y}|,
\end{equation}
where $|E_{b|y}\rangle = \mathbb{1}\otimes E_{b|y}|\mathbb{1}\rangle$ is the vectorized form of operator $E_{b|y}$, where $|\mathbb{1}\rangle$ is the unnormalized maximally entangled state (see Eq.~\eqref{choi}). Finally, the party $C$ performs a measurement and discards the state. If the POVM element corresponding to the setting $z$ and outcome $c$ is given by $E_{c|z}$, then the CJ-matrix for party $C$ is given by,
\begin{equation}\label{eq:op_C}
    M_{c|z}^{C_{1}C_{2}} = \left(E_{c|z}^{C_1}\right)^T\otimes\frac{\mathbb{1}}{2}.
\end{equation}
With these simplifications, we can define a reduced $W_r$ matrix where, 
\begin{equation}\label{eq:Wrbc}
\begin{aligned}
     p(b,c|x,y,z)=\qquad\qquad\qquad\qquad\qquad\qquad\qquad\\ \operatorname{Tr}\left[W_r^T\left(\rho_{x}^{A_{2}} \otimes |E_{b|y}\rangle\langle E_{b|y}|^{B_{1}B_2}
    \otimes \left(E_{c|z}^{C_{1}}\right)^{T}\right)\right],
    \end{aligned}
\end{equation}
and
\begin{equation}\label{eqn:red_W}
\begin{aligned}
W_r = \frac{1}{2}\operatorname{Tr}_{A_1C_2}\left[W\right].
\end{aligned}
\end{equation}
We note that for parties $B$ and $C$ the operations become  CPTP when they are summed over the measurement outcomes. To obtain the success probability in the QRAC games played by $B$ and $C$, we find the reduced statistics from the joint probability distribution $p(b,c,x,y,z)$,
\begin{equation}
    \begin{aligned}
        p_{\suc}^{B} &= \sum_{xyzc}p(b\!=\!x_y,c,x,y,z)= \frac{1}{8}\sum_{x,y}p(b\!=\!x_{y}|x,y),\\
        p_{\suc}^{C} &=\sum_{xyzb}p(b,c\!=\!x_z,x,y,z)= \frac{1}{16}\sum_{x,y,z}p(c\!=\!x_{z}|x,y,z).
    \end{aligned}
\end{equation}

\section{Semi-device-independent certification of quantum non-Markovianity}\label{sec4}
\subsection{Sequential QRAC with Markovian processes}\label{RACMark}
In a Markovian process, the noise across time steps is uncorrelated. In the process matrix formalism, such a process is written as a tensor product of channels connecting the labs (Eq. \eqref{Wmark}). Here, in a sequential QRAC game with projective measurements for $B$, we provide a bound on the success probability of $C$ and show that the bound is saturated for a noiseless Markovian process. 

A three-party noiseless Markovian process, i.e. with identity channels connecting the parties, is written as
\begin{equation}
    W = \rho^{A_{1}}\otimes|\mathbb{1}\rangle\langle\mathbb{1}|^{A_{2}B_{1}}\otimes|\mathbb{1}\rangle\langle\mathbb{1}|^{B_{2}C_{1}}\otimes\mathbb{1}^{C_{2}},
\end{equation}
where $|\mathbb{1}\rangle\langle\mathbb{1}|^{X_{2}Y_{1}}$ is the CJ matrix of the identity channel connecting the output Hilbert space of party $X$ to the input Hilbert space of party $Y$, and $\rho^{A_{1}}$ is the input state of $A$ 
(the choice of which does not affect the joint statistics we are interested in).

The linearity of the success probability in the QRAC game implies that the upper bound will be for pure state preparations of $\rho_{x_{0}x_{1}}$. In this case, we show that if we perform a rank-one projective measurement corresponding to all the settings and outcomes of $B$ and send the post-measurement state to $C$, the success probability of $C$ is upper bounded by $3/4$ (see Appendix \ref{bound_mar_appendix}). 

If the post-measurement states received by $C$ are $\tilde{\rho}_{00}$, $\tilde{\rho}_{01}$, $\tilde{\rho}_{10}$, and $\tilde{\rho}_{11}$, the optimal success probability of $C$ is given by (see Appendix \ref{bound_mar_appendix} for details)
\begin{equation}\label{succ_C_norm}
\begin{aligned}
    p_{\suc}^{C} = \frac{1}{2} + \frac{1}{8}\norm{\frac{\tilde{\rho}_{00}+\tilde{\rho}_{01}}{2} - \frac{\tilde{\rho_{11}}+\tilde{\rho}_{10}}{2}}_{1} \\+ \frac{1}{8}\norm{\frac{\tilde{\rho}_{00}+\tilde{\rho}_{10}}{2} - \frac{\tilde{\rho}_{11}+\tilde{\rho}_{01}}{2}}_{1},
\end{aligned}
\end{equation}
where $\norm{\cdot}_{1}$ is the $l_{1}$ norm. Therefore, $p_{\suc}^{C}$ can be interpreted in terms of distance between states. It is well known that under a quantum channel (CPTP map), the $l_{1}$ norm distance is non-increasing, i.e., for a channel $\mathcal{C}(\cdot)$, and states $\rho_{1}$ and $\rho_{2}$, we always have $\norm{\mathcal{C}(\rho_{1})-\mathcal{C}(\rho_{2}) }\leq \norm{\rho_{1}-\rho_{2}}$. Therefore, in the presence of a non-trivial channel from $B$ to $C$, both the distance terms in \eqref{succ_C_norm} can only decrease, leading to a reduced success probability in $C$. Analogously, a non-trivial channel from $A$ to $B$, i.e., introducing decoherence in the prepared state before it reaches party $B$ does not improve the bound on the success probability of $C$. In fact, as we show in the Appendix \ref{bound_mar_appendix}, starting with an arbitrary preparation, as long as sharp measurements are applied at $B$, the success probability of $C$ is upper bounded by the optimal classical success probability. Along with the previous argument for a non-trivial channel between $B$ and $C$, we observe that any process of the form
\begin{equation}
    W = \rho^{A_{1}}\otimes [\mathcal{T}_{1}]^{A_{2}B_{1}}\otimes[\mathcal{T}_{2}]^{B_{2}C_{1}}\otimes\mathbb{1}^{C_{2}}
\end{equation}
where, $\rho$ is a state and $[\mathcal{T}_{i}]$ are CJ matrices of the channels, i.e., for all Markovian processes, the optimal success probability for party $C$ is upper bounded by the optimal classical success probability, i.e., $p_{\suc}^{C}\leq3/4$. 

\subsection{Sequential QRAC with classical non-Markovian processes}\label{SeqRACclnonMark}
\begin{figure}[h]
    \centering
    \includegraphics[width=0.85\columnwidth]{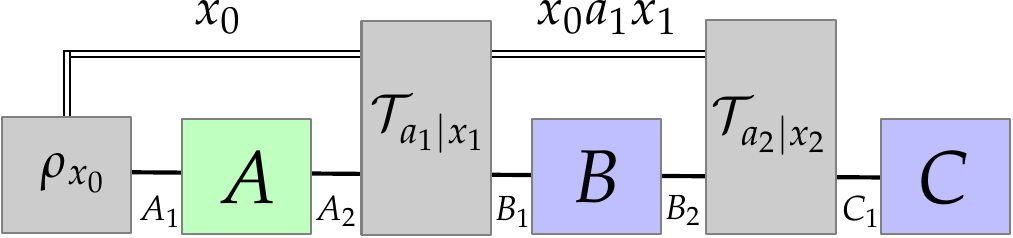}
    \caption{A general classical memory process explicitly showing the dependence on the classical variables.} 
    \label{fig:classmem}
\end{figure}

A general classical memory process for three parties has the following form (starting with Eq. \eqref{classicallong})
\begin{equation}
\begin{aligned}
       W_{CM} = \sum_{\Vec{x}\Vec{a}} &p(x_{0})p(x_{1}|x_{0})p(x_{2}|a_{1}x_{0}x_{1}) \rho_{x_{0}}^{A_{1}}\\
       &\otimes[\mathcal{T}_{a_{1}|x_{1}}]^{A_{2}B_{1}}\otimes[\mathcal{T}_{a_{2}|x_{2}}]^{B_{2}C_{1}}\otimes\mathbb{1}^{C_{2}},
\end{aligned}
\end{equation}
where $\Vec{a} = (a_1, a_2)$; $a_j$ is a classical outcome of the environment measuring the system between times;  $\rho_{x_{0}}^{A_{1}}$ is a normalized state; and $[\mathcal{T}_{a_{1}|x_{1}}]^{A_{2}B_{1}}$ is a CJ matrix of a CP trace non-increasing map $\mathcal{T}_{a_{1}|x_{1}}$ (see Fig. \ref{fig:classmem}). The relevant statistics depend only on the reduced process from Eq. \eqref{eqn:red_W} which leads to the simplified form,
\begin{equation}
    W_{CM} = \sum_{j,i} p(i) [\mathcal{T}_{j|i}]^{A_{2}B_{1}}\otimes[\mathcal{C}_{ij}]^{B_{2}C_{1}},
\end{equation}
where $\mathcal{T}_{j|i}$ is a CP map and when summed over $j$ becomes a CPTP map, and $\mathcal{C}_{ij}$ is a CPTP map $\forall$ $i,j$. Now, operations in the labs $A$, $B$, and $C$ as in \eqref{eq:op_A}, \eqref{eq:op_B} and \eqref{eq:op_C} yield the following probabilities,
\begin{equation}
\begin{aligned}
    p(b,c|x,y,z) = \qquad\qquad\qquad\qquad\qquad\qquad\qquad\\\sum_{i,j}p(i) \operatorname{Tr}\Big[E_{c|z}\mathcal{C}_{ij}\left(\mathcal{M}_{b|y}(\mathcal{T}_{j|i}(\rho_{x}))\right)\Big]
    \end{aligned}
\end{equation}
where we have used the inverse map $\operatorname{Tr}_{I}((\rho)^{T}\otimes\mathbb{1}^{O}[\mathcal{M}]^{IO}) = \mathcal{M}(\rho)$, and $\mathcal{M}_{b|y}(\cdot) = P_{b|y}(\cdot)P_{b|y}$, such that $P_{b|y}$ is a rank-one projective operator. Note that $\mathcal{T}_{j|i}(\rho_{x}) = p(j|i)\rho_{x}^{ij}$, where $p(j|i) = \operatorname{Tr}(\mathcal{T}_{j|i}(\rho_{x}))$ and $\rho_{x}^{ij}$ is a normalized state. Now, to find the bound on success probability of $C$, we need $p(c|x,z)$, which is obtained to be
\begin{equation}
    \begin{aligned}
        p(c|x,z) = &\sum_{i,j}p(ij) \operatorname{Tr}\Big[E_{c|z}\mathcal{C}_{ij}\left(\mathcal{M}^{P}( \rho_{x}^{ij})\right)\Big]\\
        =& \sum_{ij}p(ij)p_{ij}^{P}(c|x,z)
    \end{aligned}
\end{equation}
where the superscript on $\mathcal{M}^{P}$ represents projective measurements were performed in the lab $B$, and  $p_{ij}^{P}(c|x,z)$ represents the reduced statistics obtained when preparation  $\rho_{x}^{ij}$ reaches lab $B$ and statistics at $C$ given projective measurements at $B$, i.e., for each $i,j$, it represents a statistics obtained from Markovian process. Now, $p_{\suc}^{C}$ is a linear function of statistics, therefore, the resulting success probability for a general classical memory process will be a convex sum of success probability obtained through some Markovian processes with projective measurement at $B$, i.e., 
\begin{equation}
    p_{\suc}^{C}(W_{CM}) = \sum_{ij}p(ij)f_{L}(p_{ij}^{P}(c|x,z)) \leq 3/4
\end{equation}
where, we have used the result from the previous section that each term $f_{L}(p_{ij}^{P}(c|x,z))$ is upper bounded by $3/4$, where $f_{L}$ represents a linear function on $p(c|x,z)$ yielding $p_{\suc}^{C}$. Therefore, we observe that a classical non-Markovian process does not improve the bound on success probability of $C$, given projective measurements were applied in the lab $B$.

A classical non-Markovian process therefore, yields no sequential advantage, even with unconstrained success probability at $B$. Hence, a violation of this bound would semi-device-independently certify the presence of \emph{quantum} non-Markovianity in the process. We show that this is possible in the following sections.

\subsection{Feasibility problem for sequential advantage and quantum non-Markovianity}\label{Feasibilityprob}

\begin{figure}[h]
    \centering
    \includegraphics[width=0.9\columnwidth]{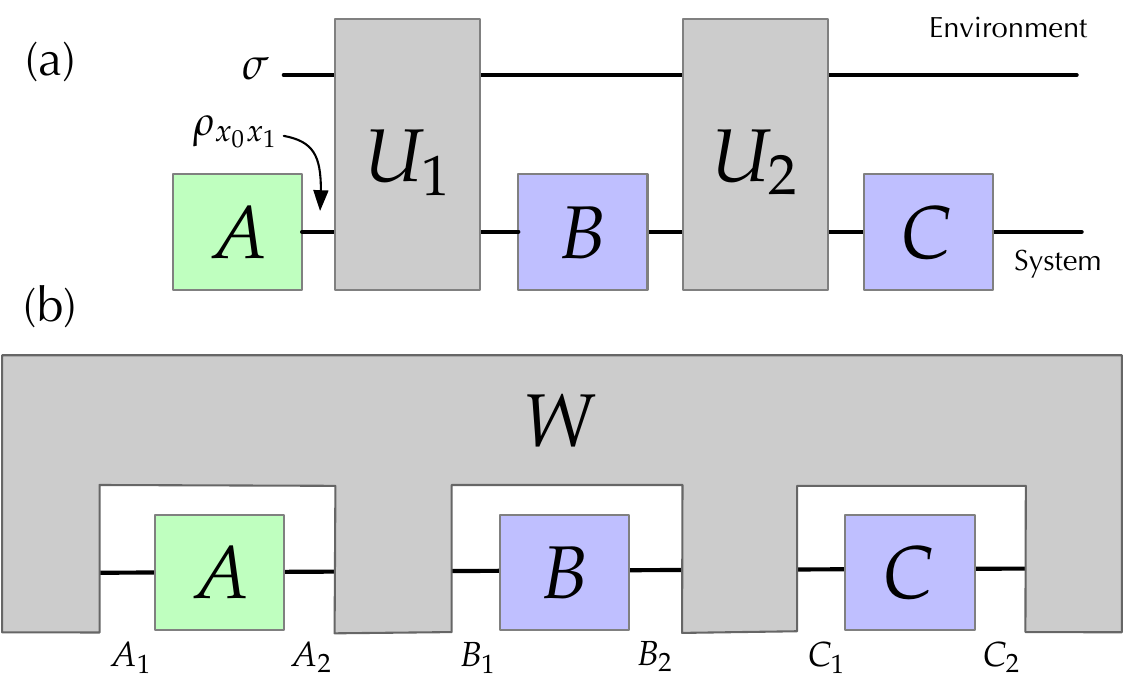}
    \caption{(a) An environment-system model for the sequential QRAC game. (b) The process matrix formulation of the game.} 
    \label{fig:envW}
\end{figure}

We now have a feasibility problem at hand --- whether there exist processes yielding sequential quantum advantage. It is a search over all possible environments and system-environment interactions. If we approach this problem traditionally, the possibility of an arbitrary dimensional environment and arbitrary interactions in different time steps make this feasibility problem highly non-linear and difficult to solve even numerically. For the scenario depicted in Fig.~\ref{fig:envW}(a), the statistics at $B$ are 
\begin{equation}
    p(b|x,y) = \operatorname{Tr}[U_{1}(\rho_{x}\otimes\sigma)U_{1}^{\dagger}P_{b|y}\otimes\mathbb{1})],
\end{equation}
where, $\sigma$ is the initial environment state, $U_{1}$ is a joint system-environment unitary and $P_{b|y}$ is the generalized measurement operator of $B$, which we assume to be rank-one projectors. The reduced statistics for $C$ in terms of the system-environment unitaries are
\begin{equation}
    p(c|x,z) = \operatorname{Tr}[(N_{c|z}\otimes\mathbb{1})(U_{2}\tilde{\rho}_{x}U_{2}^{\dagger})],
\end{equation}
with 
\begin{equation}
    \tilde{\rho}_{x} = \frac{1}{2}\sum_{b,y}(P_{b|y}\otimes\mathbb{1})U_{1}(\rho_{x}\otimes\sigma)U_{1}^{\dagger} (P_{b|y}^{\dagger}\otimes\mathbb{1}),
\end{equation}
where $U_{2}$ is the joint system-environment unitary acting between $B$ and $C$, and $N_{c|z}$ is the POVM corresponding to setting $z$ and outcome $c$ of party $C$. Hence, the feasibility problem is a search over all states $\sigma$ and joint unitaries $U_{1}$ and $U_{2}$ in $SU(d_{S}\times d_{E})$ such that both $p_{\suc}^{B}\geq 3/4$ and $p_{\suc}^{C}\geq 3/4$. The optimisation problem is
\begin{equation}
    \begin{aligned}
 &\text{max}~~p_{\suc}^{C}\\
        &s.t.~~ p_{\suc}^{B}> 0.75\\ &\qquad U_{1}, U_{2} \in SU(d_{S}\times d_{E})\\
        &\qquad\sigma\in\mathcal{L}(H^{E}), ~~\operatorname{Tr}(\sigma) = 1\\
        &\qquad N_{c|z}\geq0,~~\sum_{c}N_{c|z} = \mathbb{1},~~\forall z\\
        &\qquad P_{b|y}\in\text{Herm}(\mathcal{H}^{B}),~~\sum_{b}P_{b|y} = \mathbb{1},~~\forall y\\
    &\qquad P_{b|y}^{2} = P_{b|y}~~\forall~b,y
    \end{aligned}
\end{equation}
where $d_{S(E)}$ is the dimension of the system (environment). Note that even though the system dimension is fixed, the environment can have arbitrary dimensions. As is evident, the optimization problem in the above form is a highly non-linear problem and difficult to solve. In contrast, process matrices provide a way to approach the problem in terms of linear constraints, hence allowing for an SDP formulation, that can be solved efficiently as we discuss next. 

\subsection{Process matrix approach to feasibility problem}\label{processmatfeasible}
In this section, we reformulate the feasibility problem with process matrices as an SDP that can be solved efficiently \cite{nesterov94} .As a process matrix implicitly includes all system-environment interactions, a search for possible environments and interactions is equivalent to the search over the set of valid processes (Fig. \ref{fig:envW}(b)). In addition, a process matrix is a linear operator on the tensor product of input and output Hilbert spaces, therefore, the search is over a finite-dimensional space; 64 dimensional in our case, which reduces to 16 after discarding $A_1$ and $C_2$. A valid process $W\in A_{1}\otimes A_{2}\otimes B_{1}\otimes B_{2} \otimes C_{1}\otimes C_{2}$ satisfies the following constraints,
\begin{equation}
    \begin{aligned}
    W&\geq 0\\
    \operatorname{Tr}(W) &= d_{A_{2}}d_{B_{2}}d_{C_{2}}\\
    L_{V}(W) &= W
    \end{aligned}
\end{equation}
where,
\begin{equation}
\begin{aligned}
       L_{V}(W) 
       =& (I-\!\!\!\!\!\!\!\!\prod_{X\in\{A,B,C\}}\!\!\!\!\!\!\!\!(I-X_2+X_{12})+A_{12}B_{12}C_{12})W
\end{aligned}
\end{equation}
where $(X)W = \operatorname{Tr}_{X}(W)\otimes\mathbb{1}^{X}/d_{X}$.
To restrict the search in a subspace with a specific causal order, namely, $A\rightarrow~B\rightarrow~C$, we have the additional constraint,
\begin{align}
    L_{ABC}(W) &= W, \\
    L_{ABC}(W) &= (I-(I\!-\!A_2)B_1B_2C_1)\nonumber\\
    &\times(I-(I\!-\!B_2)C_1)C_2)W
\end{align}
Note that these linear constraints on $W$ allow for the feasibility problem to be written as an SDP. 
To search for processes yielding sequential quantum advantage, we first fix the operations of the parties, so the objective function and constraints are linear with the only variable $W$,
\begin{equation}
    \begin{aligned}
        M_{x}^{A_{1}A_{2}} &=  \mathbb{1}^{A_{1}}\otimes \rho_{x}^{A_{2}}\\
        M_{b|y}^{B_{1}B_{2}} &= {P}_{b|y}^{B_{1}}\otimes {P}_{b|y}^{B_{2}} \\
        M_{c|z}^{C_{1}C_{2}} &= E_{c|z}^{C_{1}}\otimes\mathbb{1}^{C_{2}} /2.
    \end{aligned}
\end{equation}
with $M_{x}^{A_{1}A_{2}}$, $\sum_{b}M_{b|y}^{B_{1}B_{2}} $ and $\sum_{c}M_{c|z}^{C_{1}C_{2}}$ are CJ matrices corresponding to CPTP maps for all settings $x, y$ and $z$. Here, we have considered preparation as in Eq. \eqref{stateprep}  and the projective measurement in the labs $B$ and $C$ as in Eq. \eqref{proj1}. The objective function is the success probability of $C$, as our feasibility problem is concerned with obtaining successive quantum advantages. In addition, we can set a minimal bound, $p_0$, on the success probability of $B$ through a linear constraint. The optimisation problem as an SDP is,
\begin{equation}
    \begin{aligned}
        \text{max}~~~&p^{C}_{\suc}(W)\\
        \text{s.t.}~~~&W\geq0\\
        &\operatorname{Tr}(W) = 8\\
        &L_{V}(W) = W\\
        &L_{ABC}(W) = W\\
        &p_{\suc}^{B}\geq p_{0}.
    \end{aligned}
    \label{eq:SDP}
\end{equation}
The existence of a process with sequential quantum advantage is guaranteed when the optimal value is greater than $3/4$, for $p_{0}\in [3/4,(2+\sqrt{2})/4]$. Moreover, since the operations by $B$ are projective measurements, the feasible process must be \emph{quantum} non-Markovian. We used both MatLab and Julia to run the above SDP and found that the feasibility problem is feasible i.e., there exist processes that yield sequential quantum advantage. Note that when implementing the SDP and working with the reduced process matrix $W_r$, it is only necessary to enforce the condition $L_{BC}W=W$ where $L_{BC}=I-C_1+B_2C_1$ as by construction $W_r$ has $C_2$ last and $A_1$ first. 

\begin{figure}[h]
    \centering
    \includegraphics[width=8cm]{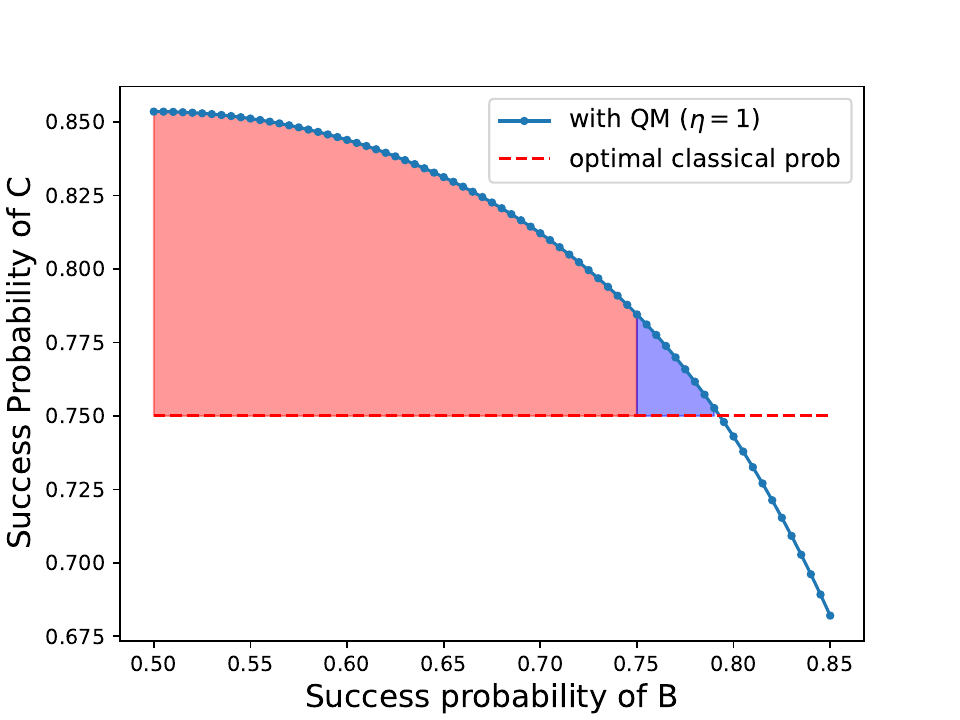}
    \caption{Optimal success probability of $C$ for various success probabilities of $B$ with sharp measurement ($\eta=1$) at $B$. The shaded regions are where one can certify quantum non-Markovianity (QNM). The blue shaded region represents process yielding sequential advantage, and in this region QNM can be robustly certified.
    }
    \label{fig:RAC2}
\end{figure}

In Fig.~\ref{fig:RAC2}, we show the optimal value of $p_{\suc}^{C}$ for various lower bounds $p_{0}$ of $p_{\suc}^{B}$ with projective measurement at the intermediate party. To compare, for the analogous noiseless, Markovian or classical non-Markovian scenarios, the success probability at $C$ is upper bounded by $3/4$. Hence, $p_\suc^C > 3/4$ allows us to infer the presence of a coherently correlated environment. If the instrument at $B$ is completely trusted, namely if $B$ deterministically performs any arbitrary projective measurement at each round of the game, a quantum advantage at $C$ would certify quantum non-Markovianity. 
The pink region in the plot is where $p_{\suc}^B \in [\frac{1}{2},\frac{3}{4}]$ and $p_{\suc}^C$ gives quantum advantage. But complete trust on the instrument makes the certification device-dependent and unreliable. In particular, there is no robustness bound on the trustability of the instrument in this region, as this joint probability can be simulated in different ways (and consequently loses its practicality). For example, $B$ performs some sub-optimal classical strategy without disturbing the system at all, and $C$ achieves quantum advantage without the presence of any kind of environment. Alternatively, $B$ can perform projective measurements in some rounds, but not all and $C$ can achieve overall quantum advantage, which can also result in such joint probabilities without the need of an environment. Therefore, to semi-device-independently certify quantum memory with \emph{minimum trust} on the instrument, successive quantum advantages for both parties are necessary.

Both parties getting quantum advantage enforces $B$ to perform some non-classical strategy, which leaves open only two possibilities that can exhibit sequential success --- the unsharpness of the instrument or quantum memory in the environment. This permits a practically feasible scenario where assuming a minimum trust on the instrument (lower bound on the unsharpness) of $B$, we can robustly certify the process having quantum memory.

In Fig.~\ref{fig:RAC2}, the blue-shaded region corresponds to a sequential quantum advantage with sharp measurement at $B$ and thus certifies quantum memory with only assumptions that Alice encoded the message in qubits, and Bob performed any arbitrary projective measurement. Having a robustness bound on the assumptions about the type of measurement at $B$ makes the certification testable in experiments, as we show in the subsection \ref{subsec:Robust}. 

Therefore, from the joint measurement statistics, we are able to certify the presence of a quantum non-Markovian environment. The $p_{\suc}^{C}$ is monotonically decreasing with $p_{\suc}^{B}$ exhibiting the trade-off relation between the two. As the success probability of $B$ increases, there is a decrease in the information loss to the environment in the first time step (since the operation in the lab $B$ is fixed). This leads to a smaller temporal correlation across time steps and consequently lower contribution of the feedback to make up for the decoherence introduced by the projective measurements at $B$; finally leading to a lower optimal success probability at $C$.

\subsection{Entanglement in processes and sequential advantage}\label{EnatnglProcess}

We have shown that with projective measurement at $B$, the quantum non-Markovian process is the only resource allowing sequential quantum advantage. Importantly, classical memory processes, where the correlation across time steps is classical, fails to yield sequential advantage and thus coherent correlations (non-classical) across time steps is necessary. We find that correlations across time steps increase monotonically with the optimal quantum advantage at $C$. 
\begin{figure}[]
    \centering
    \includegraphics[width=8cm]{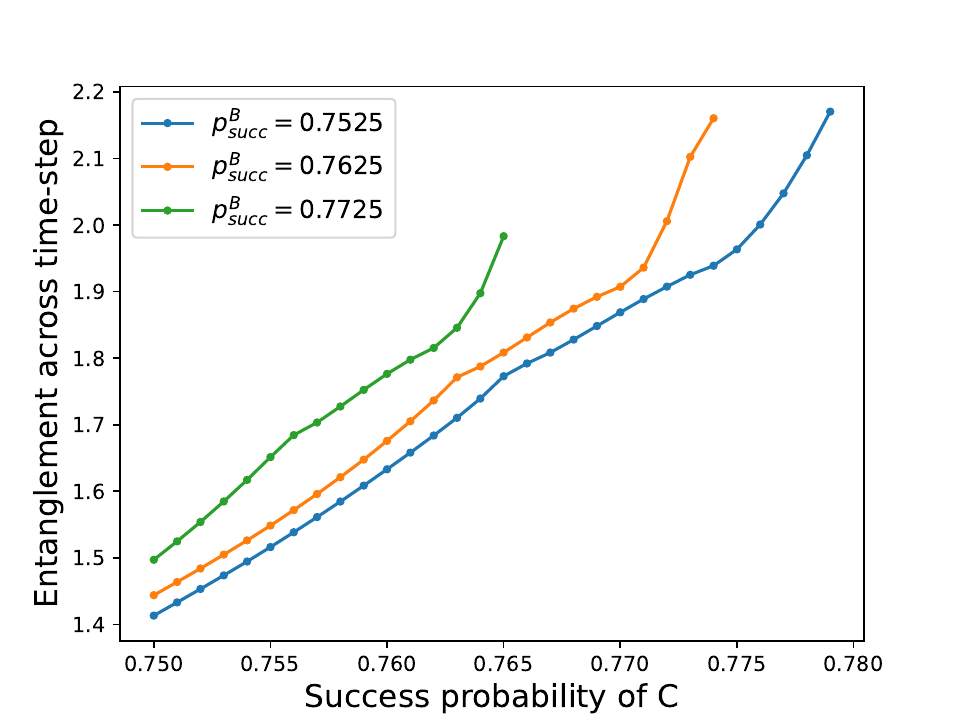}
    \caption{An increase in the quantum advantage at $C$ corresponds to higher coherent correlation across the time steps. Here, entanglement is quantified across bipartition $A_{2}B_{1}|B_{2}C_{1}$.}
    \label{fig:RAC3}
\end{figure}

In Fig.~\ref{fig:RAC3}, we plot a measure of correlations across the time steps with various success probability of $C$ for fixed success probability of $B$. As a measure of correlations we use entanglement of the bipartite state $\rho^{XY}$, where $X= A_2B_1$ and $Y=B_2C_1$, as quantified by \textit{negativity}, which is an entanglement monotone, and is quantified by the sum of magnitude of negative eigenvalues of the partially transposed state $(\rho^{XY})^{T_{X}}$ \cite{Werner_2002_PRA}. To obtain the process corresponding to the success probability $p_{\suc}^C$ for a given $p_{\suc}^{B}$, we used the SDP in Eq.~\eqref{eq:SDP} with an additional linear constraint $p_{\suc}^{W}\leq a$, where $a$ is the success probability required for $C$. The SDP provides a feasibile process for the given $(p_{\suc}^{B},p_{\suc}^{C})$, in which we investigate the bipartite correlation as described above. We observe that the amount of entanglement manifests in higher optimal sequential success probability. We emphasize that the entanglement detection in process requires process tomography, which assumes full trust in the measurement, in addition to requiring informationally complete measurements. However, in our approach, the violation of the bound on success probability of $C$ certifies the presence of quantum memory without trusting the preparation and measurements in the labs. Therefore, it is a stronger form of quantum memory certification.

\subsection{Robustness analysis: Quantum memory certification with unsharp measurements at lab $B$}\label{subsec:Robust}
\begin{figure}[t]
    \centering
    \includegraphics[width=0.5\textwidth]{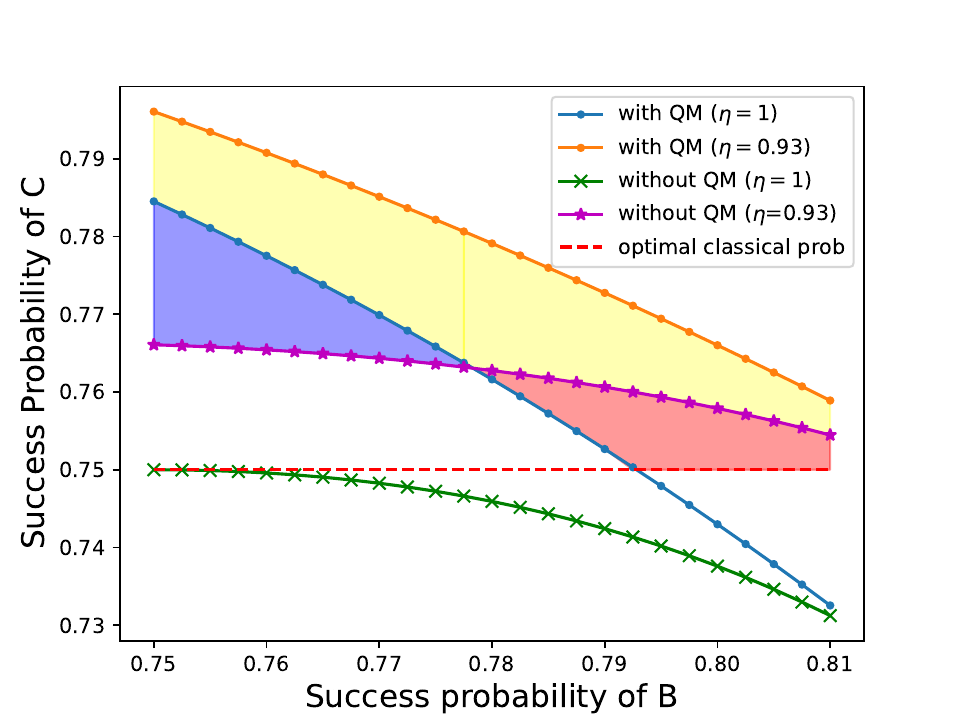}
    \caption{The blue and green plots represent the upper bound on $p^{C}_{\suc}$ for sharp measurement ($\eta=1$) with and without environment, respectively. The orange and purple plots represent the upper bound on $p^{C}_{\suc}$ for unsharp measurement (lower bound $\eta=0.93$)  with and without environment, respectively. }
    \label{fig:RAC4}
\end{figure}

No measurement is perfect in a realistic experimental set-up. Therefore, the assumption that Bob's instrument performs arbitrary projective measurements is quite restrictive. Since the optimal success probability for $C$, with projective measurements at $B$, is upper bounded by the optimal \emph{classical} success probability (plot in green in Fig.~\ref{fig:RAC4}), a very small sequential quantum advantage would be insufficient to faithfully certify quantum memory. This is because a sequential advantage could be obtained if $B$ performs \emph{unsharp} measurements, causing less disturbance to the measured state which leaves more extractable information at $C$. 
Hence, we introduce an unsharpness parameter $\eta$ on the measurement operator for $B$ \cite{Mohan_2019,PhysRevLett.125.080403}
\begin{equation}
    M_{b|y} = \sqrt{\frac{1+(-1)^{b}\eta}{2}} P_{0|y} + \sqrt{\frac{1+(-1)^{b+1}\eta}{2}} P_{1|y},
\end{equation}
where $\eta \in [0,1]$, and $P_{b|y}$ are arbitrary rank-one projective measurements. The measurement operator reduces to the identity (noninteractive) measurement for $\eta = 0$ and to sharp measurements for $\eta = 1$. 

In the noiseless scenario, the success probability of $C$ as a function of $\eta$ is (see Appendix~\ref{appendixC} for derivation),
\begin{equation}
        p_{\suc}^{C} = \sqrt{1-\eta^{2}}~p_{C}^{I} + (1-\sqrt{1-\eta^{2}})p_{C}^{S},
\end{equation}
where $p_{C}^{I}$ is the success probability of $C$ when $B$ performs the identity measurement and $p_{C}^{S}$ is the success probability of $C$ when $B$ performs rank-one projective (sharp) measurements. Note that the choice of measurements at $C$ for which $p_{C}^{I}$ and $p_{C}^{S}$ maximizes is different. As a result, to achieve a high success probability at $C$ for a given $\eta$, there is a trade-off between $p_C^I$ and $p_C^S$. 

In Fig.~\ref{fig:RAC4} we compare the optimal $p^C_\suc$ for varying  $p^B_\suc$ with and without the presence of quantum non-Markovianity as well as with and without unsharp measurements at $B$. In particular, the green plot represents optimal $p^C_\suc$, for $\eta=1$ i.e., sharp measurement at $B$ in a noiseless scenario. If Bob's instrument is unsharp, the upper bound on $p_{\suc}^{C}$ increases, as is evident in the purple plot in Fig.~\ref{fig:RAC4}, which is for a noiseless scenario with unsharpness $\eta=0.93$. The blue plot represents the optimal achievable sequential advantage with sharp measurement and quantum memory. As seen from the blue-shaded area in Fig.~\ref{fig:RAC4}, we can still certify the presence of quantum memory through the violation of the new bound assuming a lower bound on the unsharpness parameter, i.e., $\eta\geq0.93$. This gives a robustness bound to experimentally certify quantum memory with partial trust in the instrument at $B$. However, the sequential advantage in the pink-shaded region can not be achieved only through the coherent correlation in the environment. Therefore, a sequential advantage in this region certifies an unsharp instrument at $B$. Finally, the sequential advantage in the yellow-shaded region implies the presence of quantum non-Markovianity as well as unsharpness at $B$, where the upper bound has been obtained with $\eta\geq0.93$ (orange plot). Therefore, we observe that assuming a lower bound on the unsharpness parameter allows certification of either quantum non-Markovianity or unsharp instruments, or both.

\begin{figure}[t]
    \centering
    \includegraphics[width=8cm]{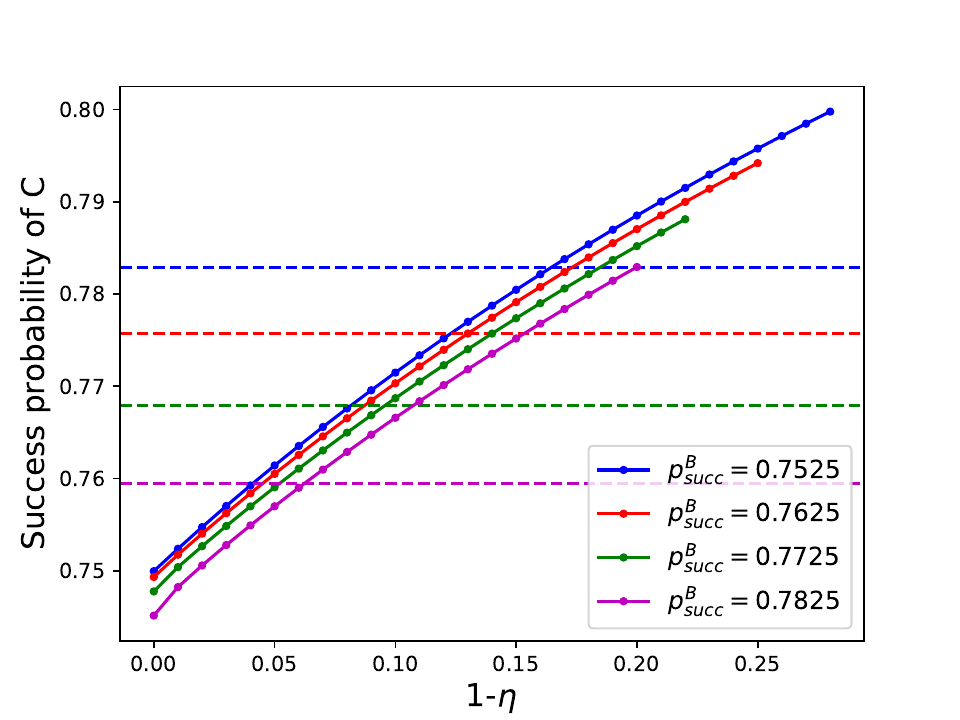}
    \caption{Upper bound on $p_{\suc}^{C}$ for different unsharpness parameter $\eta$ given a fixed values of $p_{\suc}^{B}$  for process without quantum memory. The horizontal line corresponds to the optimal feasible process with quantum memory and sharp measurement at $B$.}
    \label{fig:RAC5}
\end{figure}

Note that, whenever the optimal sequential success probability curve for quantum memory with sharp measurement is above the curve for unsharp measurement without environment, one can find a region of joint probabilities that uniquely attributes the sequential advantage to quantum memory. This is possible by assuming a lower bound on the unsharpness parameter. In Fig.~\ref{fig:RAC5}, we plot optimal $p^C_\suc$ as a function of the unsharpness parameter for various fixed $p^B_\suc$, in the absence of an environment. The horizontal dashed lines correspond to the optimal $p^C_\suc$ for given $p_{\suc}^{B}$ with a coherently correlated environment and sharp measurement at $B$. The intersection provides the upper bound on $1-\eta$, above which optimal sequential advantage from unsharpness takes over the successive advantages from quantum memory. Therefore, the regions in the plots before the intersection provide certification of quantum memory, assuming the intersection point is the minimum value of the unsharpness parameter. It is evident that the lower bound on $\eta$ allowing certification of quantum memory decreases with an increase in $p_{\suc}^{B}$, i.e. for a larger success probability of $B$, the certifiable region for quantum memory reduces. 

\begin{figure}[t]
    \centering
    \includegraphics[width=0.35\textwidth]{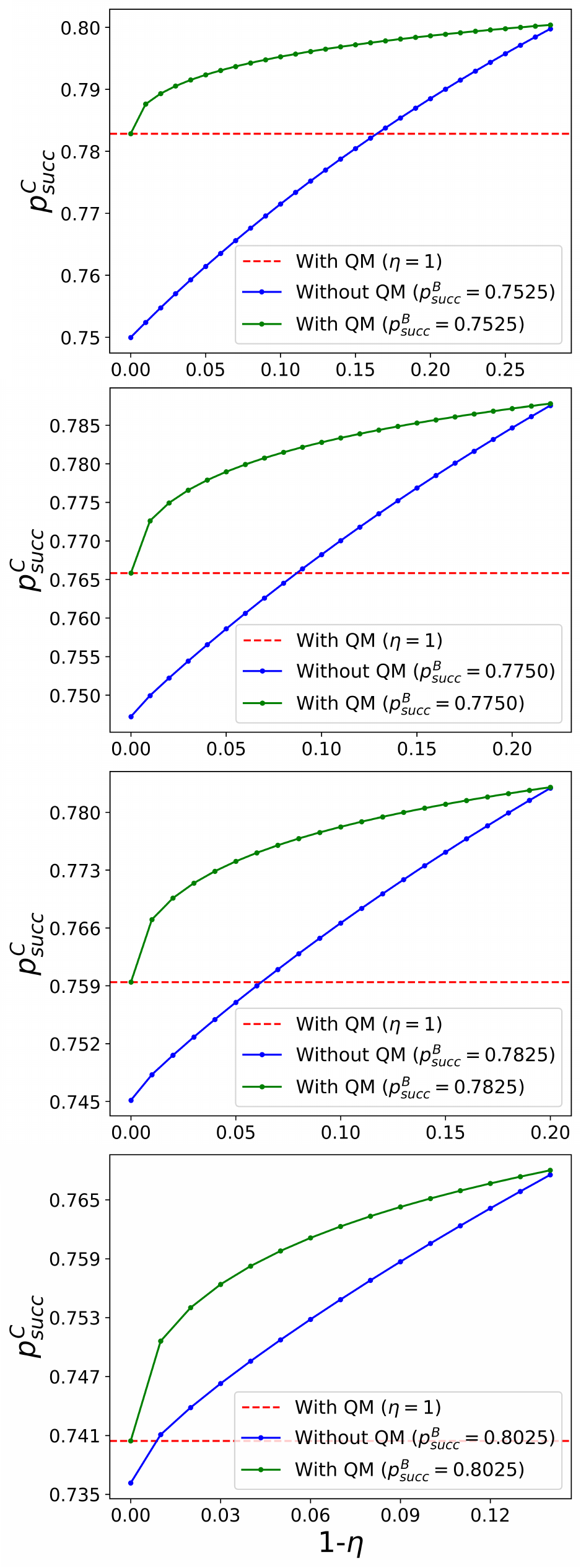}
    \caption{Optimal success probability of $C$ with (blue curve) and without (green curve) quantum non-Markovian environment for varying unsharpness parameter for fixed values of $p_{\suc}^{B}$ in the respective feasible regions. }
    \label{fig:RAC8}
\end{figure}

In Fig.~\ref{fig:RAC8} we plot optimal $p^C_\suc$ as a function of the unsharpness at $B$'s instrument in a noiseless setting (blue plots) as well as with a coherently correlated environment (green plots) for different fixed values of $p^B_\suc$ in the feasible region and compare it with the optimal case with quantum memory and sharp measurement (red dashed line in horizontal). Note that, the blue plots being below the reds gives us certifiable regions for quantum memory, while the regions after intersection certify unsharpness.  In all plots, the optimal achievable sequential success probability is higher for a correlated environment, showing that quantum non-Markovianity is a resource, even with unsharp measurements. For the minimum $\eta$ with a feasible solution for fixed $p^B_\suc$, the blue and the green plots coincide implying the optimal process is the noiseless scenario, i.e., where there are identity channels between the different labs.  

\subsection{Unambiguous Semi-device-independent Certification of Unsharp Measurement}\label{Unambiguous}

\begin{figure}[]
    \centering
    \includegraphics[width=8cm]{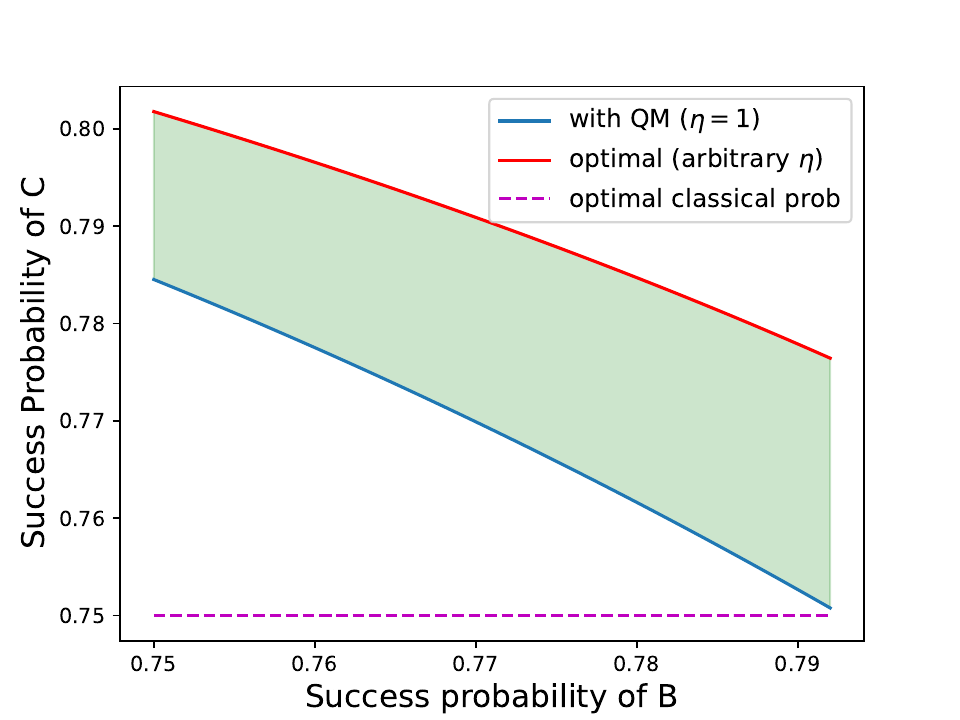}
    \caption{Optimal joint success probability for $B$ and $C$ for arbitrary unsharp measurement (red curve) as compared to the optimal success probability with quantum memory and sharp measurement at $B$ (blue curve) in the region where sequential wins are feasible. The shaded region is where one can unambiguously certify unsharpness.}
    \label{fig:RAC10}
\end{figure}

In the transformation step of standard prepare-transform-measure QRAC, only measurement-induced reduction of post-measurement states has been considered in the literature \cite{Mohan_2019,Miklin2020PRR,Foleto2020Experimental,Mukherjee2021semi,Mukherjee2023PRA,PhysRevLett.125.080403}. In a noiseless scenario, any sequential quantum advantage can certify the unsharpness of Bob's instrument. However, more general transformations include system-environment interactions that can aid in sequential quantum advantage as we have shown. To be specific, if more general transformations are considered, any sequential quantum advantage \emph{cannot unambiguously} certify unsharpness in the instrument as the advantage can come either from quantum memory or from unsharpness, or both. However, there are joint statistics that can not come only from quantum memory with sharp measurement, i.e., the instrument at $B$ must perform unsharp measurement. These joint probabilities are represented in the green shaded region in Fig.~\ref{fig:RAC10}. The red curve represents the joint optimal success probability for both observers for arbitrary unsharp measurement, while the blue curve represents quantum memory with only sharp measurement. Evidently, any success probability above the blue curve semi-device-independently certifies the presence of an unsharp instrument at $B$ in an unambiguous manner with only assumptions that qubits were prepared. Also, we note that if the unsharpness parameter is large, precisely for $\eta \geq 0.97$, the sequential quantum advantage from unsharpness never goes above the QM curve. Therefore, if the unsharpness parameter $\eta \geq 0.97$, it can never be unambiguously certified in a semi-device-independent manner without assumption on the system-environment interaction.  

\section{Discussion and Conclusions}\label{sec5}
 
While QRAC games have been traditionally studied in the noiseless scenario \cite{Tavakoli2015PRL, Mohan_2019,Miklin2020PRR,Saha2023PRA}, since the presence of environment is usually associated with detrimental decoherence effects on the system which worsens the success probabilities, here we focus on the regime of strong system-environment interactions. In particular, we show that a quantum non-Markovian environment can act as a resource to provide sequential wins at QRAC games with projective measurement at the intermediate party; something not possible in the noiseless scenario. 
We show that with assumptions on the dimension of the system and a minimum of trust in the intermediate instrument, we can unambiguously certify a quantum non-Markovian environment. Alternatively, we can view the quantum non-Markovian environment as allowing sequential wins in an QRAC game \emph{despite} having intermediate projective measurements. 

In addition, we argue that the \emph{unsharpness} of the instrument can be semi-device-independently certified for certain joint probabilities even in the presence of quantum non-Markovianity. In contrast to previous works \cite{Mohan_2019,PhysRevLett.125.080403}, we show that the more general setting allowing system-environment interactions leads to tighter bounds on the unambiguous certification of the unsharpness (see Fig.~\ref{fig:RAC10}).

In future work, it will be interesting to find the classes of system-environment interaction resulting in such resourceful non-Markovian processes. For example, a system-environment interaction model which simulates all the quantum advantage at lab $C$ when success at $B$ is unconstrained is given by partial swap operations between the labs. This model exhibits quantum memory but does not lead to \emph{sequential} quantum advantage. Though we have found processes that provide sequential advantage, they were provided \emph{numerically} through an SDP. These lack an explicit physical model for yielding successive quantum advantage. Importantly, the set of processes is a convex set, therefore with any finite set of processes yielding successive quantum advantage, one can find an infinite number of resourceful processes through convex linear combinations.

Furthermore, it is worthwhile to ask if the presence of a non-Markovian environment allows for an extension of sequential quantum advantage to more than two independent observers. Interestingly, we find that there is no process allowing sequential quantum advantage at another lab $D$, with rank-one projective measurements at intermediate labs $B$ and $C$. Even allowing unsharp measurements at intermediate labs does not permit the optimal success probability at $D$ to be more than $3/4$. Moreover, when we lift the causal restriction $L_{ABCD}(W) = W$, which corresponds to allowing indefinite causal order between labs $B$ and $C$, there is no quantum advantage at $D$. In the noiseless scenario, it has been suggested in the previous works \cite{Mohan_2019,Mukherjee2021semi}
that the sequential win in QRAC games cannot possibly be extended to another third sequential independent observer, based on its connection to the temporal Clauser–Horne–Shimony–Holt (CHSH) game. Interestingly, in a recent article \cite{Steffinlongo2022PRL} the authors show that even with sharp measurement, nonlocal correlation is extractable for sequential CHSH game for three independent observers. A rigorous theoretical investigation of sequential quantum advantage in a prepare-transform-measure scenario is left for future work. It will also be interesting to explore if using higher dimensional input and output Hilbert spaces leads to extending the sequential advantage to more than two parties. 

\section*{Acknowledgement}
SM and AKR thank Alok Kumar Pan and Asmita Kumari for initial discussions on sequential RAC. We also thank Sumit Mukherjee and Fabio Costa for useful discussions. AKR and VS acknowledge funding from Sydney Quantum Academy. SM acknowledges support from the CSIR project 09/0921(16634)/2023-EMR-I. CG acknowledges funding from the UTS Chancellor's Research Fellowship. This work was supported by the Australian Research Council (ARC) Centre of Excellence for Quantum Engineered Systems grant (CE170100009).

Finally, we acknowledge the traditional owners of the land on which the University of Technology Sydney, and Macquarie University are situated, the Gadigal people of the Eora Nation, and Wallumattagal clan of the Dharug Nation.

\bibliography{references}

\appendix
\section{Bound for Markovian processes}\label{bound_mar_appendix}

\subsection{Success probability upper bound for rank-1 projective measurement at lab B}
Here we show that even for arbitrary preparations at $A$, the optimal classical success probability bounds the success probability of $C$, when the channels connecting the labs are identity channels, and the measurements at $B$ are rank-one projective measurements. In the next subsection, we show that having noisy channels in between the labs can only make the success probability of $C$ worse. 

First, we consider a noiseless Markovian process, where the process matrix is given by,
\begin{equation}\label{process_noiseless}
    W = \frac{1}{2}\mathbb{1}^{A_{1}}\otimes [\mathcal{I}]^{A_{2}B_{1}}\otimes[\mathcal{I}]^{B_{2}C_{1}}\otimes\mathbb{1}^{C_{2}},
\end{equation}
where $[\mathcal{I}]$ represents the CJ matrix corresponding to the identity channel, which is an unnormalized maximally entangled state. Consider $A$ preparing an arbitrary state $\rho_{x_{0}x_{1}}$ corresponding to the random bit $x_{0}x_{1}$. The corresponding operation in the lab $A$ will be $M_{x_{0}x_{1}}^{A_{1}A_{2}} = \mathbb{1}\otimes\rho_{x_{0}x_{1}}$. In the lab $B$, a rank-one projective measurement is performed corresponding to each setting $y$, i.e., the operations in the lab $B$ are of the form $M_{b|y}^{B_{1}B_{2}} = P_{b|y}\otimes P_{b|y}$, where $P_{b|y}$ is a rank-one projector for all settings and outcomes $b$ and $y$. Finally, $C$ performs a measurement and discards the state, which corresponds to the operation $M_{b|y}^{C_{1}C_{2}} = E_{c|z}\otimes \mathbb{1}/2$. With these operations and the process matrix~\ref{process_noiseless}, we obtain the following probabilities,
\begin{equation}
    \begin{aligned}
        p(b,c|x_{0}x_{1},y,z) &= \operatorname{Tr}\left[W^T(M_{x_{0}x_{1}}^{A_{1}A_{2}}\otimes M_{b|y}^{B_{1}B_{2}}\otimes M_{c|z}^{C_{1}C_{2}})\right]\\
        &= \operatorname{Tr}\left[E_{c|z}(P_{b|y}\rho_{x_{0}x_{1}}P_{b|y})\right].
    \end{aligned}
\end{equation}
Now, the success probability of $C$ becomes,
\begin{equation}\label{succ-prob-C1}
    \begin{aligned}
        p_{\suc}^{C} = \frac{1}{8}\sum_{x_{0}x_{1},z}\operatorname{Tr}\left[\tilde{\rho}_{x_{0}x_{1}}E_{x_{z}|z}\right],
    \end{aligned}
\end{equation}
where, $\tilde{\rho}_{x_{0}x_{1}} =  \sum_{b,y}P_{b|y}\rho_{x_{0}x_{1}}P_{b|y}/2\equiv C_{B}(\rho_{x_{0}x_{1}})$. Since, $C_{B}(\cdot)$ and trace are both linear operations, by using the spectral decomposition of the incoming states $\rho_{x_{0}x_{1}}$, it is straightforward to see that the success probability will be upper bounded by the cases when $B$ receives pure states. Therefore, we can restrict $\rho_{x_{0}x_{1}}$ to be pure states to search for a bound on success probability at $C$. Using a similar argument, for the optimal success probability, POVM elements of $C$ must be rank one projectors. Therefore, we only need to consider the scenario where $B$ receives pure states, performs rank-one projective measurements, and sends the post-measurement state to $C$, who then performs projective measurements in his lab. We have,
\begin{equation}
    \begin{aligned}
        \rho_{x_{0}x_{1}} &= |\psi^{A}_{x_{0}x_{1}}\rangle\langle\psi^{A}_{x_{0}x_{1}}|,\\
        P_{b|y} &= |\psi^{B}_{b|y}\rangle\langle\psi^{B}_{b|y}|,\\
         E_{c|z} &= |\psi^{C}_{c|z}\rangle\langle\psi^{C}_{c|z}|.
    \end{aligned}
\end{equation}
Now, the action of $B$ results in the following post-measurement states,
\begin{equation}\label{post-measure1}
    \tilde{\rho}_{x_{0}x_{1}} = \frac{1}{2}\sum_{b,y}p_{b|y}^{x_{0}x_{1}}|\psi^{B}_{b|y}\rangle\langle\psi^{B}_{b|y}|
\end{equation}
where, $p_{b|y}^{x_{0}x_{1}} = \operatorname{Tr}[\rho_{x_{0}x_{1}}P_{b|y}]$ and forms a probability distribution for the settings $x_{0},x_{1}$ and $y$. Using~\ref{post-measure1} in~\ref{succ-prob-C1}, and after some simplification, we obtain,
\begin{equation}
    \begin{aligned}
        p_{\suc}^{C}  = \frac{1}{16}\Large[8 &+ \sum_{b,y}F^{(0)}_{b|y}|\langle\psi^{B}_{b|y}|\psi^{C}_{0|0}\rangle|^{2} \\&+ F^{(1)}_{b|y}|\langle\psi^{B}_{b|y}|\psi^{C}_{0|1}\rangle|^{2}\Large],
    \end{aligned}
\end{equation}
where,
\begin{equation}
    \begin{aligned}
        F^{(0)}_{b|y} &= p^{00}_{b|y} + p^{01}_{b|y} - p^{10}_{b|y} -p^{11}_{b|y}\\
        F^{(1)}_{b|y} &= p^{00}_{b|y} - p^{01}_{b|y} + p^{10}_{b|y} -p^{11}_{b|y}.
    \end{aligned}
\end{equation}
Using the normalisation condition $\sum_{b}|\langle\psi^{B}_{b|y}|\psi^{C}_{0|0}\rangle|^{2} = 1$, we obtain the following,
\begin{equation}
\begin{aligned}
\sum_{b,y}F^{(0)}_{b|y}|\langle\psi^{B}_{b|y}|\psi^{C}_{0|0}\rangle|^{2}+ F^{(1)}_{b|y}|\langle\psi^{B}_{b|y}|\psi^{C}_{0|1}\rangle|^{2}\\
\leq \text{max}\{F^{(0)}_{0|0},F^{(0)}_{1|0}\}+\text{max}\{F^{(1)}_{0|0},F^{(0)}_{1|0}\} \\+\text{max}\{F^{(0)}_{0|1},F^{(0)}_{1|1}\}+\text{max}\{F^{(1)}_{0|1},F^{(0)}_{1|1}\}.
\end{aligned}
\end{equation}
Now, consider the terms of following form,
\begin{equation}
    \begin{aligned}
        F^{(0)}_{b_{1}|y}+ F^{(1)}_{b_{2}|y} = &(p^{00}_{b_{1}|y}+ p^{00}_{b_{2}|y}) + (p^{01}_{b_{1}|y}- p^{01}_{b_{2}|y}) \\&+ (p^{10}_{b_{2}|y}- p^{10}_{b_{1}|y}) - (p^{11}_{b_{1}|y}+ p^{11}_{b_{2}|y}),
    \end{aligned}
\end{equation}
which for both the cases when $b_{1}=b_{2}$ and $b_{1}\neq b_{2}$ has a maximum value of $2$. Therefore, we obtain the following bound on the success probability at $C$,
\begin{equation}
    p_{\suc}^{C}\leq \frac{3}{4}.
\end{equation}
To summarize, we have shown that for arbitrary preparation of $A$, if $B$ performs rank-one projective measurement, $C$ cannot have a quantum advantage in the noiseless scenario. In the following subsection, we show that the same is true for all Markovian processes.

\subsection{Success probability bound for noisy channels}
Consider the state received by Charlie as $\rho_{x_{0}x_{1}}$, and the operation performed by Charlie corresponding to the setting $z$ and outcome $c$ as $E_{c|z}$. Then, the optimal success probability can be calculated by the following SDP,
\begin{equation}\label{red-pcsucc}
    \begin{aligned}
        \text{max}~~&p_{\suc}^{C} = \frac{1}{8}\sum_{x,z}p(c=x_{z}|x,z)\\
        &\text{s.t.}~~\sum_{i}E_{i|z} = \mathbb{1},~~z=0,1\\
&~~~~~E_{i|z}\geq 0~~\forall~~ i,z
    \end{aligned}
\end{equation}
where, $p(c|x,z) = \operatorname{Tr}(\rho_{x}E_{c|z})$. We can reduce the number of constraints in the above SDP by redefining the POVMs as,
\begin{equation}\label{red-povm}
    \begin{aligned}
        E_{0|z} = \frac{\mathbb{1}+M_{z}}{2}\\
        E_{1|z} = \frac{\mathbb{1}-M_{z}}{2},
    \end{aligned}
\end{equation}
which along with the positivity constraints $E_{i|z}\geq 0$ can be combined into the following constraints,
\begin{equation}
    -\mathbb{1}\leq M_{z}\leq \mathbb{1}~~~~z=0,1.
\end{equation}
Further, writing the objective function in terms of variables $M_{z}$, now provides the following SDP,
\begin{equation}
            \begin{aligned}
        &\text{max}~~p_{\suc}^{C}\\
        &\text{s.t.}~~    -\mathbb{1}\leq M_{z}\leq \mathbb{1}~~~~z=0,1
    \end{aligned}
\end{equation}
where, the objective function explicitly is,
\begin{equation}
    \begin{aligned}
        p_{\suc}^{C} = \frac{1}{2} &+\frac{1}{16}\operatorname{Tr}[(\rho_{00} +\rho_{01}- \rho_{11}-\rho_{10})M_{0}]\\
        &+\frac{1}{16}\operatorname{Tr}[(\rho_{00} +\rho_{10}- \rho_{11}-\rho_{01})M_{1}].
    \end{aligned}
\end{equation}
Noticing the the $l_{1}$-norm of an operator $A$ has the following SDP representation \cite{Daniel_book_IOP_2022},
\begin{equation}
\begin{aligned}
    \norm{A}_{1} = &~~\text{max}~~\operatorname{Tr}(AX)\\ 
    &\text{s.t.}~~-\mathbb{1}\leq A\leq \mathbb{1}
\end{aligned}
\end{equation}
we obtain the optimal success probability at $C$ in terms of $l_{1}$-norm as is the Eq. (\ref{succ_C_norm}). In the previous section, we showed that the success probability of $C$ in the noiseless scenario (identity channel between the labs) and projective measurement at $B$, is upper bounded by the classical optimal success probability. Using the contractivity of $l_{1}$ norm, it is straightforward to see that a noisy channel between $B$ and $C$ can only decrease the success probability at C. Further, any noisy channel between $A$ and $B$ will also decrease the success probability as it introduces decoherence in the optimally prepared state. Moreover, a channel between $A$ and $B$ cannot improve the bound as we have shown in the previous subsection that regardless of the preparation, a projective measurement restricts the success probability of $C$.   Therefore, for Markovian processes and given projective measurements, the upper bound for success probability at $C$ will be for the noiseless case (as is intuitive), which turns out to be at most the classical optimal success probability.

\section{Non-signalling cases in SDP constraints}
The constraint $L_{ABC}(W)=W$ on the process matrix does not exclude no-signalling processes and therefore we need to show that the no-signalling sets does not yield any sequential quantum advantage. We first consider the case when there is no-signalling between  all three parties $A$, $B$ and $C$. The additional constraint corresponding this this will be,
\begin{equation}
    (A_{2}B_{2}C_{2})W =W.
\end{equation}
With this constraint, the process is of the following form,
\begin{equation}
    W^{A_{1}A_{2}B_{1}B_{2}C_{1}C_{2}} = \rho^{A_{1}B_{1}C_{1}}\otimes \mathbb{1}^{A_{2}B_{2}C_{2}},
\end{equation}
where $\rho^{A_{1}B_{1}C_{1}}$ is a tripartite state satisfying $\rho^{A_{1}B_{1}C_{1}}\geq 0 $ and $\operatorname{Tr}(\rho^{A_{1}B_{1}C_{1}}) = 1$. It is straightforward to show that the resulting probabilities,
\begin{equation}
\begin{aligned}
    p(b,c|x,y,z) = \operatorname{Tr}[(&\rho^{A_{1}B_{1}C_{1}}\otimes \mathbb{1}^{A_{2}B_{2}C_{2}})^T\\&\times(M_{x}^{A_{1}A_{2}}\otimes M_{b|y}^{B_{1}B_{2}}\otimes M_{c|z}^{C_{1}C_{2}})]\\
    = ~~~~&p(b,c|y,z).
    \end{aligned}
\end{equation}
Therefore, both the success probability at $B$ and $C$ is $1/2$ --- not better than a random guess. In the case, when there is no signalling from $A$ and $B$ to $C$, the process is of the following form,
\begin{equation}\label{nosig2}
    W^{A_{1}A_{2}B_{1}B_{2}C_{1}C_{2}} = W^{A_{1}A_{2}B_{1}}\otimes\mathbb{1}^{B_{2}}\otimes\rho^{C_{1}}\otimes\mathbb{1}^{C_{2}}.
\end{equation}
For the processes of the above form, the reduced probability of $C$ is obtained as,
\begin{equation}
    \begin{aligned}
        p(c|x,y,z) &= \sum_{b}{p(b,c|x,y,z)}\\
        &= \operatorname{Tr}[\rho^{C_{1}}E_{c|z}] = p(c|z).
    \end{aligned}
\end{equation}
Therefore, again in this case, the success probability at $C$ is at best random ($p=1/2$) however, with this kind of processes, we can have quantum advantages at $B$. It is worth mentioning that one of the constraints in the SDP~\eqref{eq:SDP}, we have imposed a minimum bound on the success probability at $B$. If the bound is greater than than random guess, there must be signalling from $A$ to $B$, therefore, we just needed to eliminate for processes of the form \eqref{nosig2}.

\section {Success probability bounds for unsharp measurements at lab B}\label{appendixC}
Consider arbitrary prepared states from lab $A$, $\rho_{x_{0}x_{1}}$ entering the lab $B$. Also consider arbitrary projective measurement operations  $\{P_{0|y},P_{1|y}\}$ corresponding to setting $y=0$ and $y=1$. Now we construct the following unsharp measurement using the above sharp projections,
 \begin{equation}
     \begin{aligned}
      M_{0|0} &= \sqrt{\frac{1+\eta}{2}}P_{0|0} + \sqrt{\frac{1-\eta}{2}}P_{1|0}\\
      M_{1|0} &= \sqrt{\frac{1+\eta}{2}} P_{1|0} + \sqrt{\frac{1-\eta}{2}}P_{0|0} \\
      M_{0|1} &= \sqrt{\frac{1+\eta}{2}}P_{0|1} + \sqrt{\frac{1-\eta}{2}}P_{1|1}\\
      M_{1|1} &=\sqrt{\frac{1+\eta}{2}}P_{1|1} + \sqrt{\frac{1-\eta}{2}}P_{0|1} .
    \end{aligned}
 \end{equation}
The post-measurement states from lab $B$ i.e, $\tilde{\rho}_{x_{0}x_{1}}$ will be,
\begin{equation}
    \begin{aligned}
        \tilde{\rho}_{x_{0}x_{1}} = \rho_{x_{0}x_{1}}^{S} + \frac{\sqrt{1-\eta^{2}}}{2}\mathcal{L}(\rho_{x_{0}x_{1}})
    \end{aligned}
\end{equation}
where,
\begin{equation}
    \rho_{x_{0}x_{1}}^{S} = \frac{1}{2}\sum_{b,y}P_{b|y}\rho_{x_{0}x_{1}}P_{b|y}
\end{equation}
is the post-measured state when sharp measurements $P_{b|y}$ are performed and
\begin{equation}
\begin{aligned}
    \mathcal{L}(\rho_{x_{0}x_{1}}) = &P_{0|0}\rho_{x_{0}x_{1}}P_{1|0} +P_{1|0}\rho_{x_{0}x_{1}}P_{0|0} \\&+ P_{0|1}\rho_{x_{0}x_{1}}P_{1|1} + P_{1|1}\rho_{x_{0}x_{1}}P_{0|1}
    \end{aligned}
\end{equation}
is the correction term containing the cross terms. 

Using these post-measurement states $\tilde{\rho}_{x_{0}x_{1}}$ in Eq.~\eqref{red-pcsucc} to find the success probability of $C$, we obtain
\begin{equation}\label{succ_C_weak}
    p_{\suc}^{C} = p_{C}^{S} + \frac{\sqrt{1-\eta^{2}}}{16}\sum_{x_{0},x_{1},z}\operatorname{Tr}[E_{x_{z}|z}\mathcal{L}(\rho_{x_{0}x_{1}})]
\end{equation}
where $E_{c|z}$ are the measurement operators in the lab $C$ and $p^{S}_{C}$ is the success probability at lab $C$ if sharp measurements were applied at lab $B$ ($\eta=1$).

A slight rearrangement of $\mathcal{L}(\rho_{x_{0}x_{1}})$ gives,
\begin{equation}
    \mathcal{L}(\rho_{x_{0}x_{1}}) = 2(\rho_{x_{0}x_{1}}-\rho_{x_{0}x_{1}}^{S}),
\end{equation}
which upon substitution in Eq. (\ref{succ_C_weak}) yields,
\begin{equation}\label{unsharppsucc}
    p_{\suc}^{C} = \sqrt{1-\eta^{2}}p_{C}^{I} + (1-\sqrt{1-\eta^{2}})p_{C}^{S},
\end{equation}
where $p_{C}^{I}$ is the success probability of $C$ if the prepared state directly reached $C$ without any disturbance (or $B$ performed identity measurements). In the above form, success probability is a convex sum of success probabilities at $C$ corresponding to the two extreme choices of $B$, namely performing sharp measurements or performing trivial (identity) measurement. The probabilities $p_{C}^{I}$ and $p_{C}^{S}$ do not maximize for the same choices of measurements at $C$, therefore, there will be a non-trivial trade-off which will be dependent on the unsharpness parameter $\eta$. It is straightforward to observe that for a given $\eta$ and measurements at $B$ and $C$, $p_{\suc}^{C}$ optimizes for the optimal preparations (for example, Eq. \eqref{stateprep}). We numerically optimize Eq. \eqref{unsharppsucc} over measurements at $B$ and $C$ to find the optimal $p_{\suc}^{C}$ in the absence of environment. In the presence of a quantum non-Markovian environment, the optimization is done using SDP, starting with a specific choice of measurement operators. Note however, this is without loss of generality as the search over process implicitly optimizes over various measurements as it includes local unitary transformations.

\end{document}